\documentclass[10pt,aps,floatfix]{revtex4-1}
\usepackage{graphicx}
\usepackage[colorlinks,linkcolor=blue,citecolor=blue,urlcolor=blue]{hyperref}
\usepackage{color}
\usepackage{amsmath}
\def\be{\begin{equation}}
\def\ee{\end{equation}}
\def\bea{\begin{eqnarray}}
\def\eea{\end{eqnarray}}
\def\sp{\;\;\;,\;\;\;}
\DeclareMathOperator\atanh{atanh}
\begin{document}
\title{\bf Dressed excitations, thermodynamics and 
relaxation in the 1D XXZ Heisenberg model}

\author{ A. Pavlis$^{1,2}$, X. Zotos$^{1,2,3}$}
\affiliation{
$^1$ITCP and CCQCN, Department of Physics, University of Crete,
70013, Heraklion, Greece}
\affiliation{
$^2$Foundation for Research and Technology - Hellas, 
71110 Heraklion, Greece}
\affiliation{
$^3$Leibniz Institute for Solid State and Materials Research,
01069 Dresden, Germany}

\begin{abstract}
In this note, we discuss the low and high temperature contribution 
of Thermodynamic Bethe Ansatz (TBA) dressed excitations 
in the thermodynamics and energy-magnetization relaxation within the 
Generalized Hydrodynamics approach in the linear response regime. 
In particular, we show how the temperature dependent dispersions 
of the excitations reproduce 
well known behavior of the specific heat, magnetic susceptibility, 
spin and energy Drude weights.
In this context, we derive a further formulation of the 
Drude weights from the finite wavevector relaxation.
Furthermore, we contrast the TBA description of thermodynamics and dynamics 
in terms of a multitude of string excitations to that 
in terms of a single quasi-particle in low energy effective theories.   

\end{abstract}
\maketitle

\section{Introduction}

The framework for studying the thermodynamic properties of 
the one dimensional spin-1/2 Heisenberg model, 
in the easy-plane antiferromagnetic regime, was first set in 
a seminal paper by Takahashi and Suzuki (TS) \cite{ts} 
along the line of the Thermodynamic Bethe ansatz (TBA) 
proposed by Yang and Yang \cite{yy}. 

The characteristic of the TS formulation was the introduction of 
an intricate structure of allowed "string excitations" depending on the 
value of the anisotropy parameter. This structure   
was originally attributed to the physical requirement of normalizability 
of the corresponding Bethe ansatz wavefunctions \cite{fz,hida}
and more recently given a group theoretic interpretation \cite{group}.
The specific heat and magnetic susceptibility 
where mostly obtained by a numerical 
evaluation of the TBA nonlinear integral equations.

In this work, aiming at a physical picture of the (thermo-) dynamics,
we look at the low energy dispersions of the underlying 
string excitations where we find that they are simple expressions in terms of dressed momenta
which however are temperature dependent. 
By reformulating the expressions of the specific heat and 
magnetic susceptibility 
we point out that, in contrast to field theoretic approaches, completely 
different string excitations correspondingly contribute.

On the relaxation functions, a very interesting recent extension 
of TBA was proposed for space-time dependent densities under the name of  
Generalized Hydrodynamics approach \cite{ghd1,ghd2} (GHD). 
By this novel method the spin and thermal Drude weights 
\cite{xz,klumper18,sk2001,xzth} were recovered
as asymptotic states of a quench from an initial 
thermal/magnetization step \cite{ilievski}.
Here, using the GHD approach in the linear approximation, 
we analyze the relaxation of wavevector-$q$ dependent 
thermal/magnetization profiles. 
As a byproduct we obtain the Drude weights as integrals over frequency 
of the wavector-$q$ relaxation spectral function, 
of course closely related to linear response conductivities.

This study provides theoretical background to present and feature 
experiments in 1D quantum magnets \cite{hess},
e.g "dynamic heat transport" \cite{paul} and
"transient grating spectroscopy" experiments \cite{doppler},  
that probe the relaxation of magnetization/thermal density profiles.  

\section{TBA formulation}
The XXZ anisotropic Heisenberg Hamiltonian for a chain of $N$ sites
with periodic boundary conditions is given by,

\begin{equation}
H=\sum_{i=1}^N J (S_i^x S_{i+1}^x+S_i^y S_{i+1}^y
+\Delta S_i^z S_{i+1}^z) -hS_i^z,
\label{heis}
\end{equation}
where $S_i^a$ are spin-1/2 operators and $S_{N+1}^a=S_1^a$.
The region $0\leq \Delta \leq 1$ is commonly 
parametrized by $\Delta=\cos\theta$ and $J$ is taken as the unit of energy. 
In the following we will closely follow the 
formulation and notation by TS \cite{ts} (see Appendix A).

In the thermodynamic limit, for the simplest case of $\theta=\pi/\nu$,
the solutions of the Bethe ansatz equations are grouped into a set of 
strings of order $l_j=j$ and parity $\zeta_j=+1,~j=1,...,\nu-1$ 
and one $l_{\nu}=+1,~\zeta_{\nu}=-1$.
More generally the anisotropy parameter $\theta$ is expressed 
as a continued fraction expansion (TS) $\theta={\pi\over \nu_1+1/(\nu_2+1/\nu_3+...)}$.
The densities of excitations $\rho_j(\lambda)$ and holes $\rho_j^h(\lambda)$ ($\lambda$ is the rapidity of the excitations) are given by,
\begin{eqnarray}
a_j&=&\sigma_j(\rho_j+\rho_j^h)+\sum_k T_{jk}\circ \rho_k
\nonumber\\
a_j(\lambda)&=&\frac{\theta}{2\pi} \frac{ v_j\sin( n_j\theta ) }
{ \cosh( \theta \lambda )-v_j \cos( n_j\theta) }
\label{aj}
\end{eqnarray}
\noindent
with $a\circ b(\lambda)=\int_{-\infty}^{+\infty} 
a(\lambda-\mu)b(\mu)d{\mu}$ 
and $T_{jk}$ the phase shifts given by TS (Appendix A).
The sum over $k$ is constrained over the allowed strings, 
depending on the value of the anisotropy $\Delta$ and $\sigma_j=\zeta_j$.
Minimizing the free energy, the standard Bethe ansatz equations for
the  temperature dependent effective dispersions 
$\epsilon_j$ at temperature $T$ (inverse temperature $\beta=1/k_BT$), 
are obtained,
\begin{eqnarray}
\epsilon_j&=&\epsilon^{(0)}_j + h{l}_j+T\sum_k \sigma_ kT_{jk}\circ
\ln(1+e^{-\beta\epsilon_k})\nonumber\\
\epsilon_j^{(0)}&=&-Aa_j,~A=2\pi\frac{J\sin\theta}{\theta},~
\beta\epsilon_j=\ln \rho_j^h/\rho_j.
\label{epsilon}
\end{eqnarray}

As the bare momentum of a particle (flipped spin from the ferromagnetic state)
is given by, 
\begin{equation}
p_1^{(0)}=+i\ln \frac{\sinh\frac{\theta}{2}(\lambda+i)}
{\sinh\frac{\theta}{2}(\lambda-i)},
\end{equation}
\noindent

\noindent
and 
\begin{equation}
a_1=\frac{1}{2\pi}\frac{\partial p_1^{(0)}}{\partial {\lambda}},
\end{equation}

\noindent 
we define "bare" $p^{(0)}_j$ and "dressed" $p_j$ 
momenta \cite{ghd1}, 

\begin{align}
\frac{\partial p_j}{\partial \lambda}=2\pi\sigma_j (\rho_j+&\rho_j^h)
=2\pi\sigma_j r_j,~~~
\frac{\partial p_j^{(0)}}{\partial \lambda}=2\pi a_j, \nonumber \\
&r_j=\rho_j+\rho^h_j
\label{def1}
\end{align}
\noindent
and rewrite eq.(\ref{aj}) as,
\begin{equation}
\frac{\partial p_j}{\partial \lambda}=\frac{dp^{(0)}_j}{d\lambda}-
\sum_k \sigma_k T_{jk} \circ  n_k \frac{\partial p_k}{\partial \mu},
\label{ajr}
\end{equation}

\noindent
with $n_k =\rho_k/(\rho_k+\rho^h_k)$.

Similarly, as the bare particle energies are $\epsilon^{(0)}_j=-Aa_j$ 
we define dressed energies $E_j=-A\sigma_j (\rho_j+\rho_j^h)$ 
so that eq.(\ref{aj}) becomes,

\begin{equation}
E_j=\epsilon^{(0)}_j-\sum_k \sigma_k T_{jk}\circ \ n_k  E_k. 
\label{ej}
\end{equation}

\noindent
Notice that the eigenvalues of conserved quantities are obtained 
by \cite{korepin}:
\begin{equation}
q_m=\Big(-J\frac{\sin\theta}{\theta}\Big)^m \frac{\partial^m p^{(0)}}
{\partial \lambda^m},
\end{equation}
\noindent
where $q_0=p_1^{(0)}$ is the momentum, $q_1=\epsilon_1^{(0)}$ the energy,
$q_2=j_1^{\epsilon^{(0)}}$ the energy current eigenvalues.

With these definitions the mean value of the total energy 
is written as,

\begin{equation}
E=\sum_j\int d\lambda \rho_j(\lambda) \epsilon_j^{(0)}(\lambda)
=\sum_j \sigma_j \int \frac{d p_j}{2\pi} n_j \epsilon^{(0)}_j
\label{ave}
\end{equation}
\noindent
or, by using eqs.(\ref{aj},\ref{ej}) and a procedure 
named "dressing" \cite{ghd1}, 
\begin{equation}
E= \sum_j\int d{\lambda} \rho_j(\lambda) \epsilon_j^{(0)}(\lambda)
=\sum_j \sigma_j \int \frac{d p^{(0)}_j}{2\pi}  n_j E_j.
\label{avedr}
\end{equation}

As the energy current is also a conserved quantity, we can further 
define the mean value of energy current as,

\begin{eqnarray}
J_E &=&\sum_j \int d\lambda \rho_j j^{\epsilon^{(0)}}_j
=\sum_j \sigma_j \int \frac{dp_j}{2\pi}  n_j j^{\epsilon^{(0)}}_j,
\nonumber\\
j^{\epsilon^{(0)}}_j& =&
\Big(\frac{-A}{2\pi}\Big)\frac{\partial \epsilon^{(0)}_j}{\partial \lambda}
\label{avje}
\end{eqnarray}
\noindent
or as in eq.(\ref{avedr}),
\begin{equation}
J_E =\sum_j \sigma_j \int \frac{dp^{(0)}_j}{2\pi}  n_j j^{\epsilon}_j,~~~
j^{\epsilon}_j =
\Big(\frac{-A}{2\pi}\Big)\frac{\partial \epsilon_j}{\partial \lambda}.
\label{avjedr}
\end{equation}
\noindent
By taking the derivative of eq.(\ref{epsilon}) with respect 
to the rapidity ${\lambda}$, 
we obtain an equation for the energy currents 
with the same structure as eqs.(\ref{aj},\ref{ej}).
\begin{equation}
j^{\epsilon}_j = j^{\epsilon^{(0)}}_j
-\sum_k \sigma_k T_{jk}\circ  n_k j^{\epsilon}_k.
\label{j_de_dx}
\end{equation}

At this point, it is instructive to introduce an effective velocity 
of the excitations by \cite{fk,ghd1,ghd2},
\begin{equation}
v_j=\frac{1}{2\pi\sigma_j (\rho_j+\rho_j^h)} 
\frac{\partial \epsilon_j}{\partial \lambda}=
\frac{\partial \epsilon_j}{\partial p_j}
\label{vj}
\end{equation}
\noindent
and rewrite eq.(\ref{avjedr}) in the physical form,
\begin{equation}
J_E =\sum_j \sigma_j \int \frac{dp^{(0)}_j}{2\pi} n_j (v_j E_j).
\label{avjeve}
\end{equation}

\section{Thermodynamics in the low and high temperature limit}

Having presented the basic formalism, we proceed to the low(high) temperature analysis of the dressed excitations and their contribution to thermodynamic quantities of the system in study, in particular  the specific heat and magnetic susceptibility. Let us mention, that in this section all calculations are for zero magnetic field ($h=0$).
\\
\\
Let us begin with the free energy density $f(T)$ \cite{ts,xzth} 
\begin{eqnarray}
f&=&-T\sum_j \sigma_j \int d{\lambda} a_j \ln(1+e^{-\beta \epsilon_j})
\nonumber\\
&=&-T\sum_j \sigma_j \int \frac{dp_j^{(0)}}{2\pi}
\ln(1+e^{-\beta \epsilon_j}).
\end{eqnarray}
\noindent
Consequently, we obtain the mean energy density $\epsilon(T)$, 

\begin{equation}
\epsilon(T)=\frac{\partial}{\partial \beta} (f/T)
=\sum_j \sigma_j \int \frac{dp_j^{(0)}}{2\pi}  n_j E_j,
\end{equation}
\noindent
and by evaluating the equilibrium responses, 
$\frac{\partial r_j}{\partial \beta}\Big|_{\beta,h}, 
\frac{\partial  n_j }{\partial \beta}\Big|_{\beta,h}$ (Appendix D), 
the specific heat $c(T)$,
\begin{equation}
c=\frac{\partial \epsilon }{\partial T}
=\beta^2 \sum_j \int \frac{dp_j}{2\pi}  n_j (1- n_j ) E_j^2.
\label{spec}
\end{equation}

\noindent
Similarly, we obtain the mean magnetization density $m(T)$,

\begin{eqnarray}
m&=&\frac{\partial f}{\partial h}
=\sum_j \sigma_j \int d{\lambda} a_j  n_j Q_j
\nonumber\\
&=&\sum_j \sigma_j \int \frac{dp_j^{(0)}}{2\pi} n_j Q_j,~~ Q_j=\frac{\partial \epsilon_j}{\partial h},
\end{eqnarray}

\noindent
and the magnetic susceptibility $\chi(T)$,
\begin{equation}
\chi=\frac{\partial m}{\partial h}|_{h\rightarrow 0}
=\beta \sum_j \int \frac{dp_j}{2\pi}  n_j (1-n_j) Q_j^2.
\label{chi}
\end{equation}

\noindent
The "charges", $Q_j=\partial \epsilon_j/\partial h$ also satisfy,
\begin{equation}
Q_j =Q_j^{(0)} - \sum_k \sigma_k T_{jk}\circ  n_k Q_k,
~~Q_j^{(0)}=  {l}_j.
\label{qj}
\end{equation}

At low (zero) temperatures (Appendix B) the dispersion relations 
$\epsilon_j$ are relatively 
simple functions of the dressed momenta $p_j$, which are of the order 
of temperature $T$. At $T=0$,
\begin{eqnarray}
\epsilon_1 &=& -v \sin p_1,~~~ 0 \le p_1 < \pi
\label{spinon} 
\nonumber\\
\epsilon_j&=&0,~~~j > 1,
\end{eqnarray}
\noindent
where $v=J(\pi/2)\sin\theta/\theta$ is the spinon velocity. 
$\epsilon_1$ is the dispersion of 1-string excitations (holes in the 
"magnon" Fermi sea) and is the same as the dispersion of spinons, the elementary excitations of the model, \cite{faddeev} but with only one branch.

At $T\rightarrow 0$ and $\theta/\pi=\nu$, the string excitation 
thermal energy dispersions $\epsilon_j$ form a non-overlapping sequence, 
\begin{eqnarray}
&\epsilon_1& \simeq T g(p_1) -v \sin p_1,~~~0 \leq p_1 \leq \pi
\nonumber\\
&\epsilon_j& \simeq T\ln (j^2-1) +v|p_j|,~~~j=2,...,\nu-2,
\nonumber\\
&|p_j|& \leq p_j^{max},~~~~
p_j^{max}=\frac{T}{v}\ln \Big( \frac{(j+1)^2-1}{j^2-1} \Big),
\nonumber\\
&\epsilon_{\nu-1}&\simeq T\ln(\nu-2)+v|p_{\nu-1}|,
\nonumber\\
&|p_{\nu-1}| &< p_{\nu-1}^{max},~~~
p_{\nu-1}^{max}=\frac{T}{v}\ln \Big( \frac{\nu-1}{\nu-2} \Big),
\nonumber\\
&\epsilon_{\nu}&=-\epsilon_{\nu-1},~~ p^{max}_\nu=p^{max}_{\nu-1}.
\label{et0}
\end{eqnarray}
\noindent

Note that $\ln2\leq g(p_1)\leq\ln3$, with $g(0)=g(\pi)=\ln3$ and $g(\pi/2)\simeq \ln2$ (Appendix B). Additionally the $\epsilon_j$ do not overlap as 
$\epsilon_j(\pm p_j^{max})=\epsilon_{j+1}(0)$
(the momenta are shifted by $\pm p_j^{max}$ so that the $p_j$ 
are symmetric about zero).
At $T\rightarrow +\infty$ (Appendix C),  
\begin{eqnarray}
&\epsilon_j&\simeq T\ln((j+1)^2-1),~~j=1,...,\nu-2 \nonumber\\
&|p_j|&\leq p^{max}_j,~~
p_j^{max}=\pi{j+1\over (j+1)^2-1}  \nonumber  \\              
\nonumber\\
&\epsilon_{\nu-1}&\simeq T\ln(\nu-1),~~~
p_{\nu-1}^{max}=\frac{\pi}{2}\frac{1}{\nu-1} 
\label{etinf}
\end{eqnarray}
\noindent
and thus the top of the dispersions, $\epsilon_j(p^{max}_j)/T$, are temperature independent 
as they coincide with the $T\rightarrow 0$ ones.
The momentum space is increasing with $T$ at low temperatures, 
reaching a constant value at high temperatures. 
It is also interesting to observe that a quasi-particle dispersion of the form, 
$f(p)=T\ln g -\epsilon_p$ can be interpreted as a dispersion of holes,
with degeneracy $g$, 
\begin{equation} 
 n_h=1- n=\frac{g}{g+e^{\beta \epsilon_p}}.
\end{equation} 

Concerning the excitation energies $E_j$, at $T=0$, 
$E_1=\epsilon_1$ and $E_j=0,~j>1$.
At low $T$, we find numerically that they do not have an accurate simple form, 
$E_1\sim -v|\sin p_1|$ and $E_j,~j>1$ are of $O(T)$, 
\begin{equation}
E_j\sim -T |\sin(\frac{\pi p_j}{p_j^{max}})|,~~j=\nu-1,\nu,~~
|p_j|\leq p_j^{max}, 
\end{equation}
\noindent
where $p_j^{max}$ are the same as for the $\epsilon_j$.

At this point it is interesting to compare the commonly used 
"spinon" description 
of the specific heat as obtained from a Luttinger liquid or 
bosonization theory to the TBA description. 
In the spinon description, the elementary excitations
with dispersion $\epsilon=v|\sin p|, -\pi < p \le \pi$ contribute
at low energies to the specific heat by a 4-fold 
linear dispersion $\epsilon\sim v p$, 
$c_{spinon}\simeq \frac{\pi}{3}\frac{1}{\beta v}$. 
On the other hand the TBA 1-string excitations with dispersion 
$\epsilon_1(p_1)\simeq T\ln 3-v \sin p_1$ eq.\eqref{et0} and 2-fold 
linear spectrum at low energies, give
\begin{equation}
c_{TBA}^{(1)}\simeq \beta^2 2\int_0^{+\infty} \frac{dp_1}{2\pi} 
\frac{(vp_1)^2}{4\cosh^2\frac{\beta(T\ln3-vp_1)}{2}}
\simeq 1.234\frac{1}{\beta v} 
\end{equation}
\noindent
instead of $\pi/3\simeq1.047$. In this calculation we used the $T=0$ 
dispersion $E_1$ which overestimates $c_{TBA}^{(1)}$ as at low 
temperatures $|E_1| < v \sin p_1$.
A complete, accurate numerical evaluation of eq.(\ref{spec}) of course 
reproduces the factor $\pi/3$ and also indicates that the 
contribution of higher order strings is minimal. 
As $E_j, j>1$ is $O(T)$ we find from eqs.(\ref{spec},\ref{et0})
that $c_{TBA}^{(j)} \sim \frac{1}{\beta v} \frac{1}{j^3}$. 
Note that this attribution of the low-$T$ specific heat to a single branch of 
1-string excitations is not in agreement with 
the discussion in \cite{hatsugai}.

To evaluate the magnetic susceptibility, we first note that 
in the case of zero magnetic field ($h=0$) the 
charges $Q_j$ are temperature independent 
\cite{sk2005,ilievski,Klumper_Sakai} and 
the evaluation of (\ref{qj}) 
is particularly simple giving $Q_j=0, j=1,...,\nu-2$ and 
$Q_{\nu-1}=-Q_{\nu}=+\nu/2$ (Appendix E). 
Using the relation (\ref{chi}) for the magnetic susceptibility 
and eq.(\ref{et0}) we obtain for $T\rightarrow 0$,
\begin{equation}
\chi=\frac{1}{\pi v} K,~~~K=\frac{1}{2}\frac{1}{1-1/\nu},
\end{equation}
\noindent
where $K$ is the Luttinger liquid parameter.
In the high temperature limit, $\beta \rightarrow 0$, 
from eq.(\ref{etinf}), $\chi=\frac{\beta}{4}$.
The fact that the $l_{\nu}=+1$ excitations with parity $\zeta_{\nu}=-1$ 
account for the magnetic susceptibility is not surprising, as they  
physically correspond to a uniform change of the $S^z$ component 
of the magnetization by $\pm 1$ \cite{ovch} and detected 
e.g. in ESR experiments \cite{s1}. 

The calculation of these two thermodynamic quantities poses 
a basic question on the relation between a description in terms of a    
multitude of TBA string excitations and one in terms of a 
single quasi-particle 
with linear dispersion in low energy effective theories. 
We would have assumed that the TBA string excitations dispersions are linear in 
the dressed momenta so that the single effective quasi-particle 
represents in some sense a resummation of their contributions.
This is indeed the case for the thermal energies $\epsilon_j$ but 
not for the dressed energies $E_j$ which are given by more complex 
relations. Thus, although the exact numerical TBA calculation and the 
effective theories give the same result at low temperatures, 
their relation remains not well understod. 
Closing this section, we note that 
for $\theta/\pi=\nu_1+1/\nu_2$ (see Appendix B, C), 
relations similar to eqs.(\ref{et0},\ref{etinf}) 
give the same low/high temperature asymptotic specific heat and 
magnetic susceptibility.

\section{Energy - magnetization relaxation}

In the GHD approach, the occupations $ n_j$ depend on space and time 
following the continuity equation \cite{ghd1,ghd2},

\begin{equation}
\frac{\partial  n_j(x,t)}{\partial t}+
v_j(x,t)\frac{\partial  n_j(x,t)}{\partial x}=0,
\label{ghd}
\end{equation}
\noindent
This is conjectured to be valid in the long wavelength - time limit
(hereafter, the dependence of quantities in space-time will be 
explicitly denoted by $(x,t)$, otherwise they will refer 
to  an equilibrium state
at a temperature $k_B T=1/\beta$ and magnetic field $h$).
Notice that this relation follows by making local 
the conservation of energy for every excitation mode \cite{ghd1},
  
\begin{eqnarray}
\frac{\partial E(x,t)}{\partial t}&+&\frac{\partial J_E}{\partial x}=0
\nonumber\\
\sum_j \sigma_j \int \frac{dp^{(0)}_j}{2\pi} 
\frac{\partial }{\partial t} n_j E_j
&+&
\sum_j \sigma_j \int \frac{dp^{(0)}_j}{2\pi} 
\frac{\partial }{\partial x}  n_j (v_j E_j) =0
\nonumber\\
\frac{\partial }{\partial t} \rho_j 
&+&
\frac{\partial }{\partial x} (v_j \rho_j) =0,
\end{eqnarray}
\noindent
the original form of the GHD equation.

Most of the studies have considered a quench scenario, namely two regions at different temperatures/magnetic fields initially separated by a wall. Here, with view to future experiments on quantum magnets \cite{doppler}, we want to study the energy/ magnetization relaxation, 
starting from an initial condition where a 
small sinusoidal field of wavevector-$q$, 
$\delta \beta(x)=\delta\beta_qe^{iqx}$ or $\delta h(x)=\delta h_qe^{iqx}$ 
is applied to the system, resulting in a response with the same 
wavevector-$q$ (the discussion here is closely related to one in the 
context of the Lieb-Liniger Bose gas \cite{ghd3}). 
In the following, we will explicitly denote quantities 
depending on space-time, otherwise equilibrium ones are implied.
We will first consider 
a temperature perturbation $\delta \beta_q e^{iqx}$ around the equilibrium 
state at inverse temperature $\beta$ and magnetic field $h$, 
$ n_j(x,t)=n_j + \delta  n_j(x,t)=
 n_j + \delta  n_j (t) \delta \beta_q e^{iqx}$.
Substituting in eq.(\ref{ghd}) we obtain, 
\begin{equation}
\delta  n_j(x,t)=\frac{\partial  n_j}{\partial \beta}\Big|_{\beta,h} 
\delta\beta_q e^{iq(x-v_j t)}
\end{equation}

The space-time dependence of the energy eq.(\ref{ave}) becomes,
\begin{eqnarray}
E(x,t)&=& \sum_j \int d{\lambda} 
r_j(x,t)  n_j(x,t) \epsilon_j^{(0)}
\nonumber\\
 &=& \sum_j \int d{\lambda} 
(r_j + \delta r_j(x,t)) 
( n_j +\delta  n_j(x,t)) \epsilon_j^{(0)}
\nonumber\\
E(x,t)&=&E + \delta E(x,t)
\end{eqnarray}
\noindent
and after linearization (see Appendix D),
\begin{eqnarray}
\delta E(x,t)&\simeq& \sum_j \int d{\lambda} 
( \delta r_j(x,t)  n_j+
r_j \delta n_j (x,t) ) \epsilon_j^{(0)},
\nonumber\\
\frac{\delta E(x,t)}{ \delta \beta_q} 
&\simeq& \sum_j \int d{\lambda} 
(\frac{\partial r_j}{\partial \beta}\Big|_{\beta,h}  n_j+
r_j \frac{\partial  n_j}{\partial \beta}\Big|_{\beta,h}) 
\epsilon_j^{(0)} e^{iq(x-v_j t)},
\nonumber\\
\frac{\delta E(x,t)}{ \delta \beta_q} 
&\simeq& -\sum_j \sigma_j \int \frac{dp_j}{2\pi}  n_j(1-n_j) E_j^2 e^{iq(x- v_jt)}.
\label{deltae}
\end{eqnarray}

\noindent
Taking a Fourier transform, we obtain,

\begin{eqnarray}
\frac{1}{2\pi}\int dt e^{i\omega t} \frac{\delta E(x,t)}{ \delta \beta_q} 
&\simeq&
\nonumber\\
 -\sum_j \sigma_j \int \frac{dp_j}{2\pi} 
&n_j(1- n_j)& E_j^2 \delta(\omega-q v_j)e^{iqx}
\nonumber\\
&=& -S_{EE}(q,\omega)e^{iqx}.
\end{eqnarray}

\begin{equation}
S_{EE}(q,\omega)=
\sum_j \sigma_j \int \frac{dp_j}{2\pi} 
  n_j(1-  n_j) E_j^2 \delta(\omega-q v_j)
\label{see}
\end{equation}
\noindent
is the energy structure factor in the $q\rightarrow 0$ limit within GHD. 
In this limit, $S_{EE}(q,\omega)$ is related to the specific heat $c(T)$,
$c=\beta^2 \int d\omega S_{EE} (q,\omega)$.

In a similar analysis for the energy current $J_E$ we find that,
\begin{equation} 
\frac{\delta J_E(x,t)}{ \delta \beta_q} 
\simeq -\sum_j \sigma_j \int \frac{dp_j}{2\pi} 
  n_j(1-  n_j) E_j j_j^{\epsilon}
e^{iq(x- v_j t)}.
\label{deltaje}
\end{equation} 

\noindent
At this point it is interesting to observe that the derivative 
with respect to time of ${\delta J_E(x,t)}/{ \delta \beta_q}$ 
satisfies a continuity equation in $q$-space,

\begin{eqnarray} 
\frac{\partial}{\partial t} \frac{\delta J_E(x,t)}{ \delta \beta_q} 
&\simeq& iq\sum_j \sigma_j \int \frac{dp_j}{2\pi} 
  n_j(1-  n_j) E_j j_j^{\epsilon} v_j e^{iq(x- v_j t)}
\nonumber\\
&\simeq& iq\sum_j \sigma_j \int \frac{dp_j}{2\pi} 
  n_j(1-n_j)( j_j^{\epsilon})^2  e^{iq(x- v_jt)}
\nonumber\\
\frac{\partial}{\partial t} \frac{\delta J_E(x,t)}{ \delta \beta_q} 
&-&iq\sum_j \sigma_j \int \frac{dp_j}{2\pi} 
 n_j(1- n_j)( j_j^{\epsilon})^2  e^{iq(x- v_jt)}=0
\nonumber\\
\label{eq_dE_dt}
\end{eqnarray}
\noindent
with "current" related 
to the thermal Drude weight $D_{th}$ \cite{xzth},
\begin{eqnarray} 
D_{th}&=&\frac{\beta^2}{2}\sum_j \sigma_j \int \frac{dp_j}{2\pi} 
  n_j(1- n_j)(j_j^{\epsilon})^2
\nonumber\\
&=&\frac{\beta^2}{2}\sum_j \sigma_j \int \frac{dp_j}{2\pi} 
  n_j(1-  n_j)(v_j E_j)^2.
\label{dth}
\end{eqnarray} 

As in the case of the energy, we obtain the energy current structure factor,
\begin{equation}
S_{J_E J_E} (q,\omega)=
\sum_j \sigma_j \int \frac{dp_j}{2\pi} 
  n_j(1- n_j) (j_j^{\epsilon})^2 \delta(\omega-q v_j),
\label{sjeje}
\end{equation}
\noindent
which reduces to a $\delta$-function as $q\rightarrow 0$ with weight $D_{th}$.
In the low temperature limit $ v_j \rightarrow v$ 
and the thermal Drude weight $D_{th}=\frac{v^2}{2}c$.
Additionally, taking the second time derivative to \eqref{deltae} we obtain:
\be 
\frac{\partial^2}{\partial t^2} \frac{\delta E(x,t)}{ \delta \beta_q} 
\Big\vert_{x,t=0}
\simeq q^2 D_{th}.
\label{ballistic_E}
\ee
The above relation is interpreted as ballistic energy transport, where 
the thermal Drude weight $D_{th}$ can be seen as the inverse of an 
effective mass, $ m_{th}\sim {1\over D_{th}}$.
\\
\\
Similarly to the energy, the mean value of the magnetization $Q$ is given by,
\begin{equation}
Q=\sum_j\int d{\lambda} \rho_j(\lambda) Q_j^{(0)}
=\sum_j \sigma_j \int \frac{d p_j}{2\pi} n_j 
Q^{(0)}_j
\label{avq}
\end{equation}
\noindent
and applying a space-time dependent magnetic field 
$\delta h(x)=\delta h_qe^{iqx}$ we obtain a relaxation,

\begin{eqnarray}
\frac{\delta Q(x,t)}{ \delta h_q}
&\simeq& -\beta\sum_j \int d{\lambda}
 r_j  n_j(1-  n_j)Q_j^2 e^{iq(x- v_jt)}
\\
&\simeq& -\beta\sum_j \sigma_j \int \frac{dp_j}{2\pi}
  n_j(1- n_j) Q_j^2 e^{iq(x- v_jt)}.
\nonumber
\label{deltaq2}
\end{eqnarray}
\noindent
Finally, taking a Fourier transform, we obtain,

\begin{eqnarray}
\frac{1}{2\pi}\int dt e^{i\omega t} \frac{\delta Q(x,t)}{ \delta h_q}
&\simeq&
\nonumber\\
-\beta \sum_j \sigma_j \int \frac{dp_j}{2\pi}
 & n_j(1-  n_j)& Q_j^2 \delta(\omega-q v_j)e^{iqx}
\nonumber\\
&=&-\beta S(q,\omega)e^{iqx},
\end{eqnarray}

\noindent
with $\beta S(q,\omega)e^{iqx}$ the magnetization relaxation function
\cite{kubotomita}. $S(q,\omega)$, the $q\rightarrow 0$ spin structure factor within GHD is given by,

\begin{equation}
S(q,\omega)=
\sum_j \sigma_j \int \frac{dp_j}{2\pi}
  n_j(1-  n_j)Q_j^2 \delta(\omega-q v_j).
\label{sqq}
\end{equation}

\noindent
Again, in this limit, $S(q,\omega)$ is related 
to the magnetic susceptibility $\chi$,\;
$\chi(T)=\beta \int d\omega S(q,\omega)$.
Also, at $T\rightarrow 0$, $S(q,\omega)={K\over \beta v}\delta(\omega-v|q|)$, with $\beta v$ an effective thermal length, in contrast to the free boson model\cite{affleck_sqw}, where $S(q,\omega)=K|q|\delta(\omega-v|q|)$.
The spin current however is not a conserved quantity in the XXZ Heisenberg model, it has been conjectured though within GHD 
and recently rigorously proven \cite{klumper18} 
that its mean value is given by,
\begin{equation}
J_S =\sum_j \int d{\lambda} \rho_j (v_j Q_j^{(0)})
=\sum_j \sigma_j \int \frac{dp_j}{2\pi} n_j (v_j Q_j^{(0)}).
\label{avjs}
\end{equation}
\noindent
A similar analysis as for the energy current leads to,
\begin{align} 
&\frac{\partial}{\partial t} \frac{\delta J_S(x,t)}{ \delta h_q}- \nonumber \\
&iq\sum_j \sigma_j \int \frac{dp_j}{2\pi} 
  n_j(1- n_j)( v_j Q_j)^2  e^{iq(x- v_jt)}=0,
\end{align}
\noindent
with "current" related 
to the spin Drude weight $D_s$ \cite{xz},
\begin{equation} 
D_{s}=\frac{\beta}{2}\sum_j \sigma_j \int \frac{dp_j}{2\pi} 
 n_j(1- n_j)(v_j Q_j)^2.
\label{ds}
\end{equation} 

Moreover, the corresponding spin current structure factor is 
\begin{equation}
S_{J_S J_S}(q,\omega)=
\sum_j \sigma_j \int \frac{dp_j}{2\pi}
n_j(1-  n_j)(v_jQ_j)^2 \delta(\omega-q v_j).
\label{sqq}
\end{equation}

In the zero temperature limit, using eq.(\ref{et0}), 
$D_s$ is easily evaluated giving the known $T=0$ result \cite{ss},
\begin{equation}
D_s=\chi \frac{v^2}{2}= \frac{1}{2\pi} v K.
\end{equation}
\noindent 
The high temperature limit is particularly interesting, since (\ref{ds}) implies a "fractal" behavior \cite{prosen, affleck_ql,ilievski,klumper18} 
as a function of the anisotropy $\Delta$.
In this limit (Appendix C), for 
$\Delta=\cos \pi/(\nu_1+1/\nu_2)=\cos(\pi m/l), l=1+\nu_1\nu_2, m=\nu_2$,
\begin{eqnarray}
\epsilon_{\nu_1+\nu_2-1}&=&-\epsilon_{\nu_1+\nu_2}=
T\ln \frac{l-\nu_1}{\nu_1}
\nonumber\\
Q_{\nu_1+\nu_2-1}&=&-Q_{\nu_1+\nu_2}=\frac{l}{2}
\nonumber\\
p^{max}_{\nu_1+\nu_2-1}
&=&p^{max}_{\nu_1+\nu_2}
=\frac{\pi}{2}\frac{1}{(l-\nu_1)\nu_1}
\nonumber\\
v_{\nu_1+\nu_2-1}&=&-v_{\nu_1+\nu_2}=
\alpha\sin(\xi p_{\nu_1+\nu_2-1})
\nonumber\\
\alpha&=&\frac{\sin \pi m/l}{\sin \pi/l},\;
\xi=\frac{2\nu_1(l-\nu_1)}{l},
\end{eqnarray}
\begin{eqnarray}
&&D_s=\frac{\beta}{2}(\frac{l}{2})^2\cdot (1-\frac{\nu_1}{l})(\frac{\nu_1}{l})
\cdot \big( \sum_{j=\nu_1+\nu_2-1}^{\nu_1+\nu_2} 
\sigma_j \int_{-p_j^{max}}^{+p_j^{max}}
\frac{dp_j}{2\pi} {v_j}^2 \big)
\nonumber\\
&&D_s=\chi \frac{{\bar v}^2}{2}=\frac{\beta}{2}
\frac{\sin^2( \frac{\pi m}{l} )}{\sin^2(\frac{\pi}{l})}
(1-\frac{\sin\frac{2\pi}{l}}{\frac{2\pi}{l}}),
\nonumber\\
&&\chi=\beta/4,~~ 
{\bar v}^2= \int_{-1}^{+1} dt\alpha^2 \sin^2(\frac{\pi t}{l}),
\end{eqnarray}

\noindent
which traces the singular "fractal" behavior of $D_s$ to the velocities 
of the $\nu_1+\nu_2-1, \nu_1+\nu_2$ excitations in contrast to 
the regular behavior of $\chi$.

In analogy to \eqref{ballistic_E}, a similar relation can be derived
\be 
\frac{\partial^2}{\partial t^2} \frac{\delta Q(x,t)}{ \delta h_q} 
\Big\vert_{x,t=0}
\simeq q^2 D_{s}
\ee

In conclusion, in low energy effective theories one quasi-particle 
with linear dispersion and effective velocity \cite{affleck_sqw, sirker} 
accounts for both the specific heat, magnetic susceptibility and dynamic 
structure factors. In contrast, in the $\nu$ TBA string excitations, 
the 1-strings mostly contribute to the specific heat, 
while the $\nu-1,\nu$ to the magnetic susceptibility and corresponding dynamic structure factors.
As all string excitations have the same characteristic velocity 
at low energies, effective field 
theories seem as a re-summation of the string contributions, but further 
work is necessary to reconcile the two pictures.

\acknowledgements{
This work was supported by the Deutsche
Forschungsgemeinschaft through Grant HE3439/13 and the Onassis foundation.
 A.P. acknowledges helpful discussions with N. Papanicolaou, P. Lambropoulos and C. Psaroudaki.
X.Z. acknowledges fruitful discussions with H. Tsunetsugu, B. Doyon,
A. Kl\"umper, P. van Loosdrecht, C. Hess, A. Chernyshev, 
the hospitality of the
Institute for Solid State Physics - U. Tokyo and the University of Irvine.}

\widetext

\section{Appendix}
\subsection{Appendix: Takahashi-Suzuki formulation}

Following the TS formulation
the pseudomomenta $k_\alpha$ characterizing the Bethe ansatz
wavefunctions are expressed in terms of the rapidities ${\lambda}_\alpha$,
\begin{equation}
\cot(\frac{k_{\alpha}}{2})=\cot(\frac{\theta}{2})
\tanh(\frac{\theta  {\lambda}_{\alpha}}{2}).
\label{param}
\end{equation}
For $M$ down spins and $N-M$ up spins the energy $E$ and momentum $K$ are given by:
\begin{equation}
E=J\sum_{\alpha=1}^M (\cos k_{\alpha}-\Delta),~~~~
K=\sum_{\alpha=1}^M k_{\alpha}.
\label{enem}
\end{equation}
Imposing periodic boundary conditions on the Bethe ansatz wavefunctions
the following relations on the allowed values of the rapidities are obtained,
\begin{equation}
{\Big\lbrace} \frac{\sinh \frac{1}{2}\theta({\lambda}_\alpha+i)}
{\sinh\frac{1}{2}\theta({\lambda}_\alpha-i)}{\Big\rbrace}^N
=
\prod_{\beta=1}^M
{\Big\lbrace} \frac{\sinh \frac{1}{2}\theta({\lambda}_\alpha-{\lambda}_\beta+2i)}
{\sinh\frac{1}{2}\theta({\lambda}_\alpha-{\lambda}_\beta-2i)}{\Big\rbrace},
~~~\alpha=1,2,...M.
\label{pbce}
\end{equation}
In the thermodynamic limit, the solutions of equations (\ref{pbce}) are 
grouped into
strings of order $l_j, j=1,...,\nu$ and parity $\zeta_j=+1~ {\rm or}~ -1$. 
In the case $\theta=\pi/\nu$ the allowed
strings are of order ${l}_j=j$, for $j=1,...,\nu-1$ and parity $\zeta_j=+1$ 
of the form,
\begin{equation}
{\lambda}_{\alpha,+}^{{l},k}={\lambda}_{\alpha}^{l}+({l}+1-2k)i+O(e^{-\delta N});~~~k=1,2,...{l},
\label{s1}
\end{equation}
and strings of order ${l}_{\nu}=1$ and parity $\zeta_{\nu}=-1$,
\begin{equation}
{\lambda}_{\alpha,-}={\lambda}_{\alpha}+i\nu+O(e^{-\delta N}),~~~\delta > 0. 
\label{s2}
\end{equation}

Multiplying the terms in equation (\ref{pbce}) corresponding to
different members of a string and taking the logarithm we obtain 
the relations,
\begin{equation}
Nt_j({\lambda}_{\alpha}^j)=2\pi I_{\alpha}^j+\sum_{k=1}^{\infty}\sum_{\beta=1}^{M_k}
\Theta_{jk}({\lambda}_{\alpha}^j-{\lambda}_{\beta}^k),~~~\alpha=1,2,...M_j,
\label{basic}
\end{equation}
where $I_{\alpha}^j$ are integers (or half-integers) and $M_k$ is the number
of strings of type $k$,
\begin{eqnarray}
t_j({\lambda})&=&f({\lambda};{l}_j,\zeta_j),\label{Theta_1} \\
\Theta_{jk}({\lambda})&=&f({\lambda};|{l}_j-{l}_k|,\zeta_j\zeta_k)+
f({\lambda};{l}_j+{l}_k,\zeta_j\zeta_k)+ 
2\sum_{i=1}^{\min({l}_j,{l}_k)-1}f({\lambda};|{l}_j-l_k|+2i,\zeta_j\zeta_k), 
\label{Theta_2}\\
f({\lambda};l,\zeta)&=&2\zeta\tan^{-1}\Big[\cot({l\theta\over 2})^\zeta\tanh({\theta {\lambda}\over 2})\Big],
\label{Theta_3}
\end{eqnarray}

\noindent
and 
\be
T_{jk}(\lambda)\equiv{1\over 2\pi}{d\Theta_{jk}(\lambda)\over d\lambda}\sp \alpha_j(\lambda)\equiv {1\over 2\pi} {dt_j(\lambda)\over d\lambda}.
\label{Theta_T relation}
\ee

Additionally, we mention that in the more general case of ${\pi/\theta}=\nu_1+{1/\nu_2}$ ($\nu_2>1$), with a total number of $\nu_1+\nu_2$ string species, we have:

\be
l_j=\begin{cases}
j & 1\leq j \leq \nu_1-1 \\
1+(j-\nu_1)\nu_1 & \nu_1\leq j\leq \nu_1+\nu_2-1,~~~
\zeta_j=\exp i\pi [\frac{l_j-1}{p_0}]\\
\nu_1 & j=\nu_1+\nu_2, ~~~~~~~~~~~~~~~\zeta_{\nu_1}=-1.
\end{cases}
\label{nj_fract}
\ee

\be
\sigma_j=\begin{cases}
	1 & 1\leq j \leq \nu_1-1,~j=\nu_1+\nu_2 \\
	-1 & \nu_1\leq j\leq \nu_1+\nu_2-1
\end{cases}
\label{lj_fract}
\ee
 
For later use, the more general,$\;\nu_2>1$, case of eq.\eqref{epsilon} is presented in the form of a recursive relation, which at $h=0$ is:

\begin{align}
&\ln(1+e^{\beta \epsilon_0})=-{2\pi \beta J\sin\theta\over \theta }\delta(\lambda),\nonumber\\ 
&\beta\epsilon_{j}=s_1\circ\ln(1+e^{\beta\epsilon_{j-1}})+s_1\circ\ln(1+e^{\beta\epsilon_{j+1}})\sp j=1,...,\nu_1-2,\; j\neq 
\nu_1+\nu_2-2 \nonumber\\ 
&\beta \epsilon_{\nu_1-1}=s_1\circ\ln(1+e^{\beta\epsilon_{\nu_1-2}})+d_1\circ\ln(1+e^{\beta\epsilon_{\nu_1-1}})+s_2\circ\ln(1+e^{\beta\epsilon_{\nu_1}})\sp \nu_1,\nu_2\geq 2 \nonumber\\
&\beta\epsilon_{j}=(1-2\delta_{\nu_1,j}) s_2\circ\ln(1+e^{\beta\epsilon_{j-1}})+s_2\circ\ln(1+e^{\beta\epsilon_{j+1}})\sp j=\nu_1,...,\nu_1+\nu_2-3,\\
&\beta\epsilon_{\nu_1+\nu_2-2}=(1-2\delta_{\nu_2,2})s_2\circ\ln(1+e^{\beta\epsilon_{\nu_1+\nu_2-3}})+s_2\circ\ln(1+2e^{\beta\epsilon_{\nu_1+\nu_2-1}}+e^{2\beta\epsilon_{\nu_1+\nu_2-1}}) \nonumber \\
&\beta\epsilon_{\nu_1+\nu_2-1}=s_2\circ\ln(1+e^{\beta\epsilon_{\nu_1+\nu_2-2}}), \nonumber \\
&\beta\epsilon_{\nu_1+\nu_2}=-\beta\epsilon_{\nu_1+\nu_2-1}
\label{recursive_epsilon}
\end{align}

where $s_i,d_i$, $\; i=1,2$ are given by (TS) with $\int_{-\infty}^{\infty} s_i d{\lambda}=\int_{-\infty}^{\infty} d_i d{\lambda}={1\over 2}$\cite{ts}.

\subsection{Appendix: Dynamics in the $T\rightarrow 0$ limit}

First consider the $T=0$ case. We can easily see that
\begin{equation}
E_1 = \epsilon_1,~~~E_{j>1} = 0
\end{equation}

\begin{equation}
 n_1=\frac{1}{1+e^{\beta\epsilon_1}}=1,
\end{equation}
where we have used that $\epsilon_1<0$.
\\
\\
Therefore, using eqs.(\ref{def1},\ref{ajr}) and the Fourier transform we obtain :

\begin{equation}
\frac{\partial p_1}{\partial \lambda}=
\frac{\partial p_1^{(0)}}{\partial \lambda}- T_{11} \circ
\frac{\partial p_1}{\partial \lambda}(\omega)\Rightarrow \frac{\partial p_1}{\partial \lambda}(\omega)=
\frac{\partial p_1^{(0)}}{\partial \lambda}(\omega)-  T_{11}(\omega)\cdot
\frac{\partial p_1}{\partial \lambda}(\omega).
\end{equation}

Additionally we have that:

\begin{equation}
\frac{\partial p_1^{(0)}}{\partial \lambda}(\omega)=2\pi a_1(\omega),~~~
1+ T_{11} (\omega)=2\cosh(\omega)\cdot a_1(\omega)
\end{equation}

\begin{equation}
 \frac{\partial p_1}{\partial \lambda }(\omega)=\frac{\pi}{\cosh(\omega)}\Rightarrow
\frac{\partial p_1}{\partial \lambda}=\frac{\pi/2}{\cosh(\pi {\lambda}/2)}
\end{equation}
\begin{equation}
p_1({\lambda})=\tan^{-1}( \sinh\frac{\pi \lambda}{2} ),~~~ -\pi/2 \leq p_1({\lambda}) \leq +\pi/2
\end{equation}

Hence,
\begin{eqnarray}
E_1&=&(-J \frac{\sin\theta}{\theta} ) \frac{\partial p_1}{\partial \lambda}
=- v \cdot \sin p_1,~~~
v=J\frac{\pi}{2}\cdot \frac{\sin \theta}{\theta}\sp 0\leq p_1\leq \pi
\end{eqnarray}

\begin{equation}
v_1=\frac{\partial \epsilon_1/\partial \lambda}{\partial p_1 /\partial \lambda}
=-v \cdot \cos p_1,~~0\leq p_1\leq\pi.
\end{equation}

Therefore, we obtain that at zero temperature, $\epsilon_1=E_1$ is a spinon excitation as presented in eq.\eqref{spinon}.
\\
\\
Let us continue with the $T\rightarrow 0$ limit. In this case in order to have a consistent set of equations we need to  include terms of $\mathcal{O}(T)$ order. In fact the spinon excitation acquires a part which comes from the scattering with higher string species. Therefore the $\epsilon_1$ excitation can be written as
\be 
\epsilon_1(\lambda)=-{v\over \cosh({\pi \lambda\over 2})}+g(\lambda)T.
\ee

The exact calculation of $g({\lambda})$ is difficult, but we can instead calculate the asymptotic and ${\lambda}=0$ values of $g({\lambda})$
\\
\\
If we assume that all physical quantities of interest reach a fixed value at the limit, $\lambda\rightarrow\pm\infty$ then the convolution $A\circ B(\lambda)\vert_{\lambda\rightarrow\pm\infty}$, where $A(\lambda)$ vanishes rapidly at infinity can be written as:

\begin{equation} 
A\circ B(\lambda)|_{{\lambda}\rightarrow\pm\infty}=B_{\pm\infty}\int_{-\infty}^{\infty} A(\mu) d\mu
\label{conv_infinity}  , 
\end{equation}

where $B_{\pm\infty}\equiv\lim_{\lambda\rightarrow\pm\infty} B(\mu)$.

Therefore we can transform a set of non-linear integral equations such as eqs.(\ref{epsilon},\ref{recursive_epsilon})  into a set of algebraic equations. 
\\
\\
As a simple example let us consider the simplest non-trivial case, $\nu=3$
\\
\\
Using eqs.(\ref{epsilon},\ref{recursive_epsilon}) and the fact that $\int_{-\infty}^{\infty} T_{ij}(\mu)d\mu={1\over \pi}\Theta_{ij}(\infty)$ we obtain:

\bea
&g_{\infty}=-{1\over 3}g_{\infty}+{1\over 3}\ln(1+e^{g_{\infty}})+{2\over 3}\ln2+{2\over 3}\ln(1+\cosh\beta\epsilon_2), \\
&\beta\epsilon_2={1\over 2}\ln(1+e^{g_{\infty}}).
\label{g_inf}
\eea

Substituting $g_{\infty}=\ln\alpha$, $\alpha\neq 1$ we obtain the following equation:
\be 
w^3-6w^2+9w-4=0,
\ee
where we have defined that $w\equiv \alpha+1$.
\\
\\
Hence, the approved solution is $w=4$. Additionally, due to symmetry we get the following result :
\be 
\lim_{{\lambda} \to \pm\infty} g(\lambda)=\ln 3.
\ee

Also, we find that :
\be 
\lim_{\lambda \to \pm\infty} \epsilon_2=T\ln 2.
\ee

The above results can also be derived in closed form, for the more general case $\pi/\theta=\nu_1+1/\nu_2$,\;$\nu_2>1$. To this end, we take the asymptotic limit to eq.\eqref{recursive_epsilon}, and using eq.\eqref{conv_infinity} we reduce the system into a difference  equation for $\beta\epsilon_j$. Actually the difference equation, and of course the solution, is the same as the one we obtain to $\mathcal{O}(T)$ order by taking the high temperature limit $T\rightarrow \infty$. The physical explanation for this coincidence is that in the high temperature limit, the system excites large values of the rapidity $\lambda$, i.e. $\lambda\rightarrow\infty$, and thus to dominant order the two limits are equivalent. Therefore, the asymptotic solutions for the $\epsilon_j$ excitations are given by\cite{ts} 

\begin{align} 
&\lim_{ \lambda \to \pm\infty}{\epsilon_{j}}=T\ln\Big((j+1)^2-1\Big)\sp j=1,...,\nu-2 \nonumber\\
&\lim_{\lambda\to\pm\infty}\epsilon_{\nu-1}=T\ln(\nu-1)\sp\epsilon_{\nu}=-\epsilon_{\nu-1}
\label{high_temp_en}
\end{align}

in the case $\nu_1=\nu-1$,\;$\nu_2=1$ and
\begin{align} 
&\lim_{ \lambda \to \pm  \infty}\epsilon_{j}=T\ln\Big((j+1)^2-1\Big)\sp j=1,...,\nu_1-1 \nonumber\\
&\lim_{\lambda \to \pm  \infty}\epsilon_{j}=T\ln\Big(({1+(j-\nu_1)\nu_1+\nu_1\over \nu_1})^2-1\Big)\sp j=\nu_1,...,\nu_1+\nu_2-2 \nonumber\\
&\lim_{\lambda \to \pm \infty}\epsilon_{\nu_1+\nu_2-1}=T\ln({\nu_1\nu_2-\nu_1+1\over \nu_1})\sp\epsilon_{\nu_1+\nu_2}=-\epsilon_{\nu_1+\nu_2-1}
\label{high_temp_en_fra}
\end{align}
when $\nu_2>1$.
\\
\\
Let us follow the same procedure in order to calculate the $g(0)$ value which is also important for the low temperature dynamics of the excitations. This time however, our result will clearly no longer be exact, but due to the fact that at $T\rightarrow 0$ the quantities $\ln(1+e^{\beta \epsilon_j})$ are slowly varying around $\lambda=0$ and the rapidly vanishing form of the functions $s_i,d_i$ we expect this to be a valid approximation
\\
\\
Similarly with the previous case let us consider the simplest case, $\nu=3$.
\\
\\
Since $\epsilon_1(0)=-v+g(0)T$, eq.\eqref{recursive_epsilon} gives that $\beta\epsilon_2\simeq 0$, when $T\rightarrow 0$. Additionally, eq.\eqref{epsilon} gives 

\be
{4\over 3} g(0)= {2\over 3} \ln 2+{2\over 3}\ln(1+\cosh(\beta \varepsilon_{2}))
\ee

which yields the solution $g(0)=\ln 2$. Similarly we can find a closed form that holds for all $\pi/\theta=\nu>3$. Nevertheless, note that with increasing $\nu$ the numerical value will slightly differ from our analytical approximation. Using the same procedure that we used in the asymptotic case in order to arrive at a difference equation and the fact that this time $\epsilon_1(0)=-v+g(0)T$ and  $e^{\beta\epsilon_1(0)}\simeq 0$, at $T\rightarrow 0$, we obtain that the $\lambda=0$ values of the $\epsilon_{j}$ excitations are the following: 
\begin{align} 
&{\epsilon_1}(0)\simeq-v+T\ln2 \nonumber\\ 
&{\epsilon_{j}}(0)=T\ln\Big(j^2-1\Big)\sp j=2,...,\nu-2 \nonumber\\
&\epsilon_{\nu-1}(0)=T\ln(\nu-2)\sp\epsilon_{\nu}=-\epsilon_{\nu-1},
\label{zero_temp_0}
\end{align}

Note that a similar but more complicated relation can be constructed for the fractal case ($\nu_2>1$). 
\\
\\ 
Next let us consider the $\epsilon_j$ excitations as functions of $p_j$. To this end, we start from the definition of the excitation velocities:
\be 
{\partial \epsilon_j\over \partial p_j}(\lambda)=v_j(\lambda) 
\ee

In order to continue we assume that $ v_j$,\; $j=2,...,\nu$ converge very quickly to their asymptotic values. This  is plausible since $s_1(\lambda)$ vanishes rapidly, i.e. for $|\lambda|>\delta$, $s_1(\lambda)$ can be considered negligible, with $\delta>0$. The above is also justified by numerical calculations. In addition using eqs.(\ref{ajr},\ref{j_de_dx}) we find that $|v_j|=v$,\;$j=2,...,\nu$, for $|\lambda|>\delta_j$, where the numerical values of  $\delta_j$ are close to $\delta$. Moreover from equations (\ref{epsilon},\ref{ajr},\ref{j_de_dx}) we notice that ${\partial \epsilon_j\over \partial \lambda}$ is an antisymmetric function, while ${\partial p_j\over \partial \lambda}$ and $\epsilon_j$ are symmetric functions. Therefore for rapidities $|{\lambda}|>\delta_j$ we obtain that:
\be 
|{\partial \epsilon_j\over \partial \lambda}|=v {\partial p_j\over \partial \lambda}.
\label{e_p_v}
\ee

Since $\epsilon_{j}$ is a symmetric function we obtain that: 
\be
\epsilon_{j}=v|p_j-p_j(0)|+\Delta_j\sp j\geq 2,
\label{epsilon-pj}
\ee 

At this point it is important to mention that eq.\eqref{epsilon-pj} is not true for a small interval around $p_j(0)$, due to the fact that eq.\eqref{e_p_v} does not hold. This is expected since $\epsilon_j$ are  everywhere differentiable. Nevertheless, it is supported by the numerics that $p_j(|{\lambda}|<\delta_j)$ is approximately constant, and thus this interval is indeed a narrow one around $p_j(0)$.

Substituting eqs.(\ref{high_temp_en},\ref{zero_temp_0}) into eq.\eqref{epsilon-pj} it yields:

\begin{eqnarray}
&\epsilon_j& \simeq T\ln (j^2-1) +v|p_j|,~~~j=2,...,\nu-2,
\nonumber\\
&|p_j| &\leq p_j^{max},~~~
p_j^{max}=\frac{T}{v}\ln \Big( \frac{(j+1)^2-1}{j^2-1} \Big),
\nonumber\\
&\epsilon_{\nu-1}&=T\ln(\nu-2)+v|p_{\nu-1}|,
\nonumber\\
&|p_{\nu-1}|&\leq p_{\nu-1}^{max},~~~
p_{\nu-1}^{max}\simeq\frac{T}{v}\ln \Big( \frac{\nu-1}{\nu-2} \Big),
\nonumber\\
&\epsilon_{\nu}&=-\epsilon_{\nu-1},
\end{eqnarray}

where $p_j$ is shifted such that $\epsilon_j$ are symmetric about zero and $p^{max}_j={1\over 2}\lim_{{\lambda} \to \infty}|p_j( \lambda)|=|p_j(0)|$.

\subsection{Appendix: Dynamics in the $\beta\rightarrow 0$ limit }

First, we consider the non-fractal case, $\nu_2=1$. To begin with, in Appendix B we discussed that the dominant term of the high temperature behavior of $\epsilon_j$ coincides with the asymptotic limit, ${\lambda}\rightarrow\infty$, given by \eqref{high_temp_en}. The first correction to this behavior can be found by following the method applied in Appendix C by Takahashi and Suzuki\cite{ts}. Hence, using eq.\eqref{recursive_epsilon} the first order correction, $\mathcal{O}(1)$, for zero magnetic field $h=0$ is given by:

\begin{align}
&\epsilon^{(1)}_{j}({\lambda})=-{A\over 2(j+1)}(1+e^{-\beta \epsilon_{j}})[(j+2)\alpha_j(\lambda)-j\alpha_{j+2}(\lambda)]~~ j=1,...,\nu-2 \nonumber \\
&\epsilon^{(1)}_{\nu-1}(\lambda)=-{A\over 2}(1+e^{-\beta \epsilon_{\nu-1}})\alpha_{\nu-1}(\lambda),~~ \epsilon^{(1)}_\nu=-\epsilon^{(1)}_{\nu-1}.
\end{align}

Consequently, the dominant contribution of ${\partial \epsilon_{j}\over\partial \lambda}$ is given by:
\be 
{\partial \epsilon_{j}\over \partial \lambda}= {\partial \epsilon^{(1)}_j\over \partial \lambda},
\label{dedx}
\ee
\\
\\
A careful examination of eq.\eqref{ajr} and eq.\eqref{j_de_dx} shows that the set of integral equations corresponding to ${\partial \epsilon_{j}\over \partial \lambda}$ and $ {\partial p_{j}\over \partial \lambda}$ are essentially the same with the substitution of the driving term, $-A {\partial\alpha_j\over \partial\lambda}$ to $2\pi\alpha_j$. Hence, the dominant term of $ {\partial p_{j}\over \partial \lambda}$ will be given by:
\begin{align}
&{\partial p_{j}\over \partial \lambda}={\pi\over (j+1)}(1+e^{-\beta \epsilon_{j}})[(j+2)\alpha_j-j\alpha_{j+2}]~~ j=1,...,\nu-2 \nonumber \\
&{\partial p_{\nu-1}\over \partial \lambda}=\pi(1+e^{-\beta \epsilon_{\nu-1}})\alpha_{\nu-1},~~ {\partial p_{\nu}\over \partial \lambda}=-{\partial p_{\nu-1}\over \partial \lambda}
\end{align}

Therefore using \eqref{Theta_T relation} the momentum $p_j$ is given by:
\begin{align}  
&p_j(\lambda)={1\over 2(j+1)} (1+e^{-\beta \epsilon_{j}})\Big[(j+2)\Big(f(\lambda;{l}_j,\zeta_j)+f^\infty_j\Big)-j\Big( f(\lambda;{l}_{j+2},\zeta_{j+2})+f^\infty_{j+2})\Big)\Big],~ j=1,...,\nu-2 \nonumber \\
&p_{\nu-1}(\lambda)={1\over 2} (1+e^{-\beta \epsilon_{\nu-1}})\Big[f(\lambda;{l}_j,\zeta_j)+f^\infty_{\nu-1}\Big],~~~ p_\nu=-p_{\nu-1}
\label{p_j_1}
\end{align}
where $f(\lambda;{l}_j,\zeta_j)$ is given by \eqref{Theta_3} and ${f^\infty_j}=\lim_{\lambda\rightarrow\infty}f(\lambda;{l}_j,\zeta_j)$.
\\
\\
 $p^{max}_j=|p_j(0)|$ is given by
\begin{align}
&p^{max}_j=\pi{j+1\over (j+1)^2-1},\; j=1,...,\nu-2  \nonumber  \\              
&p^{max}_{\nu-1}={\pi\over 2}{1\over \nu-1}\sp p^{max}_\nu=p^{max}_{\nu-1}
\end{align}

The $\nu-1$ excitation velocity is given by:
\be 
v_{\nu-1}=-{A\over 2\pi} {\partial \alpha_{\nu-1}\over \partial{\lambda}}{1\over \alpha_{\nu-1}}.
\label{vel_1}
\ee

Let us rewrite $\theta {\lambda}$ in terms of the momentum $p_{\nu-1}$ 
\be 
\theta {\lambda}=2 \atanh\Big( \tan({\nu-1\pi\over 2\nu})\cdot\tan({\nu-1\over \nu}p_{\nu-1})\Big),
\ee
where we have shifted the momenta $p_{\nu-1}$ by $p^{max}_{\nu-1}$.
\\
\\
If $p_{\nu-1}\simeq 0$ we obtain that
\be 
\theta \lambda\simeq  2{\nu-1\over \nu}\tan({\nu-1\pi\over 2\nu})p_{\nu-1}.
\label{p_x_linear}
\ee

Therefore in this limit 
\be 
v_{\nu-1}\simeq {\sin\theta\over 1+\cos\theta} \theta \lambda=\tan{\pi\over 2\nu}\cdot\theta \lambda
\label{v_linear_x}
\ee

Substituting eq.\eqref{p_x_linear} into eq.\eqref{v_linear_x} we obtain
\be 
v_{\nu-1}\simeq 2{\nu-1\over \nu}p_{\nu-1}.
\label{v_linear_p}
\ee

On the other hand, asymptotically $|v_{\nu-1}|=\sin\theta$. Therefore, a suitable function that satisfies both regions is 
\be 
v_{\nu-1}=\sin (2{\nu-1\over \nu}p_{\nu-1})
\ee

By directly plotting $v_j(\lambda)$ as a function of $p_{\nu-1}(\lambda)$  we find that $v_{\nu-1}$ is indeed described by the above form.
\\
\\
Next we move to the $\nu_2>1$ case. Initially, let us mention that the dominant term of the excitation energies $\epsilon_j$ is given by eq.\eqref{high_temp_en_fra}. In this case we prove that the excitation velocity $v_{\nu_1+\nu_2-1}$ appears to have a fractal behavior consistent with the findings for the spin Drude weight $D_s$ at high temperatures. A simple generalization of the previous case shows that ${\partial\epsilon_{j}\over \partial \lambda}$ and momentum ${\partial p_j\over \partial \lambda}$ are given by:

\begin{align}
&{\partial\epsilon_{j}\over \partial \lambda}=-{A\over 2(n_j+1)}(1+e^{-\beta \epsilon_{j}})\Big[(n_j+2){\partial\alpha_j\over \partial\lambda}-n_j{\partial\alpha_{j+2}\over \partial\lambda}\Big],~~ j=1,...,\nu_1-1 \nonumber \\
&{\partial\epsilon_{j}\over \partial \lambda}=-{A\over 2\nu_1(n_j+\nu_1)}(1+e^{-\beta \epsilon_{j}})\Big[(n_j+2\nu_1){\partial\alpha_j\over \partial\lambda}-n_j{\partial\alpha_{j+2}\over \partial\lambda}\Big],~~ j=\nu_1,...,\nu_1+\nu_2-2 \nonumber \\
&{\partial\epsilon_{\nu_1+\nu_2-1}\over \partial \lambda}=-{A\over 2\nu_1}(1+e^{-\beta \epsilon_{\nu_1+\nu_2-1}}){\partial\alpha_j\over \partial\lambda},~~ {\partial\epsilon_{\nu_1+\nu_2}\over \partial \lambda}=-{\partial\epsilon_{\nu_1+\nu_2-1}\over \partial \lambda}
\end{align}

\begin{align}
&{\partial p_{j}\over \partial\lambda}={\pi\over (n_j+1)}(1+e^{-\beta \epsilon_{j}})\Big[(n_j+2)\alpha_j-n_j\alpha_{j+2}\Big],~~ j=1,...,\nu_1-1 \nonumber \\
&{\partial p_{j}\over \partial \lambda}={\pi\over \nu_1(n_j+\nu_1)}(1+e^{-\beta \epsilon_{j}})\Big[(n_j+2\nu_1)\alpha_j-n_j\alpha_{j+2}\Big],~~ j=\nu_1,...,\nu_1+\nu_2-2 \nonumber \\
&{\partial p_{\nu_1+\nu_2-1}\over \partial \lambda}={\pi\over \nu_1}(1+e^{-\beta \epsilon_{\nu_1+\nu_2-1}})\alpha_{\nu_1+\nu_2-1},~~ {\partial p_{\nu_1+\nu_2}\over \partial \lambda}=-{\partial p_{\nu_1+\nu_2-1}\over \partial \lambda},
\end{align}

where $n_j$ are given by eq.\eqref{nj_fract} 
\\
\\
Furthermore a careful calculation shows that $p^{max}_{\nu_1+\nu_2-1}$ is  
\be 
p^{max}_{\nu_1+\nu_2-1}={1\over 2\nu_1} (1+e^{-\beta \epsilon_{\nu_1+\nu_2-1}})|f^\infty_{\nu_1+\nu_2-1}|={\pi\over 2} {1\over {(1+\nu_1\nu_2-\nu_1)\nu_1}}.
\ee
Note that we have used the fact that $\zeta_{\nu_1+\nu_2-1}=(-1)^{\nu_2}$. 
\\
\\
Performing a linear approximation the velocity and momentum can be written as:

\be
v_{\nu_1+\nu_2-1}={\sin\theta\over 1+\cos{\theta\over \nu_2}} \theta {\lambda}
\label{linear_v_frac}
\ee

\be 
p_{\nu_1+\nu_2-1}\simeq  {1\over \xi} \zeta_{\nu_1+\nu_2-1} \cot\Big({1+(\nu_2-1)\nu_1\over 2}\theta\Big)^{\zeta_{\nu_1+\nu_2-1}} \theta {\lambda},
\label{linear_p_frac}
\ee
where we have substituted eq.\eqref{high_temp_en_fra} into eq.\eqref{vel_1} and $\xi= 2\nu_1{(1+\nu_1\nu_2-\nu_1)\over {1+\nu_1\nu_2}}$.
\\
\\
Therefore using \eqref{linear_p_frac} and \eqref{linear_v_frac} we conclude that:

\be
v_{\nu_1+\nu_2-1}=\zeta_{\nu_1+\nu_2-1}{\sin\theta\over 1+\cos{\theta\over\nu_2}} \tan\Big({1+(\nu_2-1)\nu_1\over 2}\theta\Big)^{\zeta_{\nu_1+\nu_2-1}} {\xi p_{\nu_1+\nu_2-1}}
\ee

One can easily prove that: 

\be 
\zeta_{\nu_1+\nu_2-1}{\sin\theta\over 1+\cos{\theta\over m}} \tan\Big({l-\nu_1\over 2}\theta\Big)^{\zeta_{\nu_1+\nu_2-1}}=-{ \sin m\pi/l\over \sin\pi/l},
\ee

where $m=\nu_2$ and $l=1+\nu_1\nu_2$.

\be 
v_{\nu_1+\nu_2-1}\simeq -{ \sin m\pi/l\over \sin\pi/l} \xi p_{\nu_1+\nu_2-1},
\ee

which proves the fractal behavior of the velocity. Moreover, asymptotically $|v_{\nu_1+\nu_2-1}|=\sin m\pi/l$. Plotting $v_{\nu_1+\nu_2-1}$ as a function of $p_{\nu_1+\nu_2-1}$ reveals that it can be described as 
\be 
v_{\nu_1+\nu+2-1}= -{ \sin m\pi/l\over \sin\pi/l}\sin\xi p_{\nu_1+\nu_2-1}
\ee

\subsection{Appendix: Proof of relation (34)}

To arrive at eq.\eqref{deltae}, it is convenient to use a matrix notation for the integral over rapidity ${\lambda}$, sum over string index $j$ and employ standard manipulations \cite{ghd1,ghd3}, using the convention that $[a]$ is a vector column, $tr[A]=\sum_j \int d\lambda A_j(\lambda)$ and $[T][A]=\sum_k (T_{jk}\circ A_k) $
\begin{equation}
\frac{\delta E(x,t)}{ \delta \beta_q} 
=tr \Big\lbrace
[\epsilon^{(0)} e^{iq(x-vt)}] [\frac{\partial r}{\partial \beta}  n] +
[\epsilon^{(0)}e^{iq(x-vt)}] [\frac{\partial  n}{\partial \beta} r]\Big\rbrace , 
\end{equation}
\noindent
solve eq.(\ref{aj}) for $[ n\frac{\partial r}{\partial \beta}]$,
\begin{eqnarray}
&&[r] = [ \lambda a ] -  [ \lambda T ] [nr]
\nonumber\\
&&[  n \frac{\partial r}{\partial \beta} ] = - [\lambda  n T] 
( [  n \frac{\partial r}{\partial \beta} ] + 
[r \frac{\partial  n}{\partial \beta}])
\nonumber\\
&&[ n \frac{\partial r}{\partial \beta}]= - 
[I+\lambda  n T]^{-1} [\lambda  n T]
[r \frac{\partial  n}{\partial \beta}] 
\end{eqnarray}
\noindent
solve eq.(\ref{ej}),

\begin{equation}
[E]=[I+\lambda  n T]^{-1} [\epsilon^{(0)}]
\end{equation}
and from eq.(\ref{epsilon}) and eq.(\ref{ej}) we find that
${\partial (\beta \epsilon_j)}/{\partial \beta}=E_j$.

\subsection{Appendix: Charges $Q_j$}

We can readily show this in the simple case $\theta=\pi/\nu$ 
by using the  recursion relations of TS for the phase shifts $T_{jk}$,
and rewriting eq.(\ref{qj}) as ($Q_0= n_0=0)$, 
\begin{eqnarray}
Q_j&=&s_1*Q_{j-1}(1- n_{j-1})+s_1*Q_{j+1}(1- n_{j+1})\nonumber\\ 
&+&\delta_{\nu-2,j}s_1*Q_\nu n_\nu,~~~ 1\leq j\leq\nu-2\nonumber\\
Q_{\nu-1}&=&\frac{\nu}{2}+s_1*Q_{\nu-2}(1- n_{\nu-2})
\nonumber\\
Q_{\nu-1}-Q_\nu&=&\nu.
\label{recursion}
\end{eqnarray}
\noindent
For $h=0$, $n_{\nu-1}+ n_{\nu}=1$ from (\ref{epsilon}), so  
we can eliminate the $Q_j,  n_j,~~j=\nu-1,\nu$ from (\ref{recursion})
obtaining a homogeneous system of equations for $Q_j,  n_j,~~j=1,\nu-2$
with solution $Q_j=0,~j=1,...,\nu-2$. 
An algebraic approach of this result was given in \cite{ilievski,klumper18}.
When the magnetic field is nonzero ($h\ne0$) the $Q_j$'s are in general 
functions of the rapidity $\lambda$. 

\bigskip
For completeness, we also study the case  
$\pi/\theta=\nu_1+1/\nu_2~(\nu_2 >1)$ 

Using that $\sigma_j$ is positive 
in the region $1\leq j\leq \nu_1-1,j=\nu_1+\nu_2$, the recursion relations 
eq.(\ref{qj}) take the form,
\begin{eqnarray}
Q_j&=&s_1*Q_{j-1}(1- n_{j-1})+s_1*Q_{j+1}(1- n_{j+1}),~~~1\leq j\leq \nu_1-2
\nonumber\\
Q_{\nu_1-1}&=&s_1*Q_{\nu_1-2}(1- n_{\nu_1-2})+
d_1*Q_{\nu_1-1}(1- n_{\nu_1-1})\nonumber\\
&-&s_2*Q_{\nu_1}(1- n_{\nu_1})
\nonumber\\
Q_j&=&s_2*Q_{j-1}(1- n_{j-1})+s_2*Q_{j+1}(1- n_{j+1})\nonumber\\
&+&\delta_{\nu_1+\nu_2-2,j}s_2*Q_{\nu_1+\nu_2} n_{\nu_1+\nu_2},
~~~\nu_1\leq j\leq \nu_1+\nu_2-2\nonumber\\
Q_{\nu_1+\nu_2-1}&=&-\frac{1+\nu_1\nu_2}{2}
+s_2*Q_{\nu_1+\nu_2-2}(1- n_{\nu_1+\nu_2-2}),~~~j= \nu_1+\nu_2-1
\nonumber\\
&&Q_{\nu_1+\nu_2}-Q_{\nu_1+\nu_2-1}={1+\nu_1\nu_2}.
\label{qj2}
\end{eqnarray}
Similarly as before, in the zero field case ($h=0$) 
the solution of eq.(\ref{qj2}) is 
$Q_j=0$ for $1\leq j\leq\nu_1+\nu_2-2$

\end{document}